\documentclass[preprint,amsmath,amssymb,aps,floatfix]{revtex4-1}
\usepackage{graphicx}
\usepackage{graphics}
\usepackage{adjustbox} 
\usepackage{dcolumn}
\usepackage{bm}
\usepackage{dsfont}
\usepackage{amsthm}
\usepackage[dvipsnames]{xcolor}
\usepackage{chngcntr}
\usepackage{apptools}
\AtAppendix{\counterwithin{lemma}{section}}
\usepackage{xy}
\usepackage{qcircuit}
\usepackage{tikz}
\usepackage{mathtools}
\usepackage{floatrow} 
\DeclareFloatFont{tiny}{\tiny}
\usepackage[capitalise]{cleveref}
\usepackage[T1]{fontenc}
\crefname{figure}{Fig.}{Fig.}

\theoremstyle{definition}

\theoremstyle{definition}

\newcommand{\tildetimes}{\mathbin{\tilde{\times}}}

\newcommand{\ads}{{\text{AdS}_3}}

\newcommand{\ds}{{\text{dS}_4}}

\newcommand{\noads}{{\text{NO-AdS}_3}}

\newcommand{\onesphere}{{\text{S}^1}}
\newcommand{\twosphere}{{\text{S}^2}}

\newcommand{\noadstwosphere}{\text{NO-AdS}_3 \times \twosphere}

\newcommand{\nodstimesR}{\text{BG-dS}_4 \times\R}
\newcommand{\reals}{{\mathbb{R}}}

\newcommand{\tr}{{\text{tr}}}
\newcommand{\dd}{{\text d}}

\newcommand{\be}{\begin{equation}}
\newcommand{\ee}{\end{equation}}
\newcommand{\bp}{\begin{pmatrix}}
\newcommand{\ep}{\end{pmatrix}}
\newcommand{\ben}{\begin{enumerate}}
\newcommand{\een}{\end{enumerate}}

\newcommand{\RP}[1]{{\mathbb{RP}^{#1}}}
\newcommand{\R}{\mathbb{R}}
\newcommand{\RR}[2]{\mathbb{R}^{#1,#2}}

\begin{document}

\title{Conformal Vacua in Boundary-Glued AdS$_3$}

\author{Lucas K. Kovalsky}
\affiliation{Quantum Algorithms and Applications Collaboratory, Sandia National Laboratories, Livermore, CA, USA
}%
\author{Shivesh Pathak}
\affiliation{Quantum Algorithms and Applications Collaboratory, Sandia National Laboratories, Livermore, CA, USA
}%
\author{Kyle Ritchie}
\affiliation{Leinweber Institute for Theoretical Physics and Department of Physics, University of California, Berkeley, California 94720, USA
}%
\affiliation{Quantum Algorithms and Applications Collaboratory, Sandia National Laboratories, Livermore, CA, USA}

\begin{abstract}
  We argue that the purely topological gluing of AdS$_3$ spacetimes, which are otherwise independent due to Dirichlet boundary conditions, can still produce non-trivial spaces after a quotient. In this direction, we extend the study of a recently discovered non-orientable AdS$_3$, which is a $\mathbb Z_2$ quotient of two AdS$_3$ spaces glued along their conformal boundaries. We find that this space is compact and boundaryless and that the quotient sends the gluing surface to a torus comprised of a null lattice with halved periods. However, the bulk wraps around both its sides, which means that as an embedded space this gluing surface forms a one-sided torus. We discover that it supports a permutation-twisted sector of two CFT$_2$s with shared Dirichlet boundary conditions and conjecture that conformal maps can be extended across this gluing surface. Using this extension, we show that the vacuum of pairs of Rindler wedges, which share a boundary in this non-orientable AdS$_3\times$S$^2$ space, are conformal to static patch vacua, which are supported on two dS$_4\times \mathbb R$ glued together at their conformal boundaries.
\end{abstract}
\maketitle

\tableofcontents
\newpage

\section{Introduction}\label{intro}

\noindent The AdS/CFT correspondence has served as a primary tool for understanding quantum gravity in asymptotically Anti-de Sitter spacetimes \cite{Maldacena:1997re}. Being the maximally symmetric spacetime with negative cosmological constant, $\Lambda<0$, AdS comes endowed with a timelike conformal boundary which naturally admits a boundary Hamiltonian, and to which gravitational observables can be dressed. Moreover, its asymptotic symmetry group is precisely the gauge group of conformal field theory. The AdS/CFT correspondence provides the precise relationship between operators in AdS and those in the CFT. Indeed, it is widely believed that quantum gravity in AdS has a formal UV completion in the form of its CFT dual. However, the problem of quantum gravity beyond asymptotically AdS spacetimes remains elusive.

One such avenue concerns understanding how AdS/CFT generalizes to cases where the asymptotic boundary is coupled to an external system, such a non-gravitating bath, or a second AdS universe. Such cases arise naturally in the context of evaporating AdS black holes ~\cite{Penington:2019npb,Almheiri:2019psf}, in double holography~\cite{Fujita_2011,Kawamoto:2023wzj}, or in the study of defect or interface CFTs~\cite{DeWolfe:2001pq,Karch:2000ct,Aharony:2006hz}. It has been suggested that boundary-glued AdS spacetimes arise, for example, in the context of double trace deformations~\cite{Kiritsis:2006hy,Aharony:2006hz}. More recently, the holography of two AdS spaces glued across a shared dynamical brane was studied in~\cite{Kawamoto:2023wzj} using the framework of BCFT, and the entanglement entropy of spaces glued topologically along their conformal boundary was studied in~\cite{Arias:2025vjy}.

AdS spaces glued along their conformal boundary cease to have a proper topological boundary, since the gluing surface becomes ``internal'', producing a single compact boundaryless (but still locally AdS) spacetime. These topological features are more typical of $\Lambda >0$ cosmological spacetimes, suggesting a potential link to closed universe holography. However, the potential link is complicated by the fact that holography is best understood with Dirichlet conditions imposed at the conformal boundary. Gluing together AdS spacetimes at their boundary, however, seems to require Neumann or mixed boundary conditions~\cite{Kawamoto:2023wzj}, which introduces dynamical gravity on the boundary, and complicates the holographic analysis.  This raises the question of whether AdS spaces can be meaningfully glued while maintaining the simplicity ensured by standard Dirichlet boundary conditions.

In what follows, we proceed with the aim of understanding boundary-glued AdS spaces and their relationship to closed universe holography by investigating the conformal relationship between patches of a novel, non-orientable asymptotically AdS spacetime with Dirichlet boundary conditions, and static patches of a asymptotically dS spacetime that is glued to itself.  In~\cite{Pathak24} it was demonstrated that a non-orientable AdS spacetime ($\noads$) can be constructed via an orientation-reversing $\mathbb Z_2$ quotient of two AdS spacetimes glued along their conformal boundaries. We reinvestigate this space, and find that this non-orientable AdS space contains a single timelike asymptotic internal boundary with a torus topology constructed from a null-periodic lattice and with a vacuum structure that differs from that of ordinary AdS/CFT. We argue that the boundary supports entangled states of two permutation-twisted CFTs, each dual to bulk states supported on opposite sides of this internal boundary, where the bulk wraps around. Furthermore, as a consequence of the null boundary topology, the two CFTs globally factorize into chiral and antichiral components.

We then conjecture that a such internal conformal boundaries allow a conformal analytic extension across themselves, in a higher-dimensional generalization of the Schwarz reflection principle. Given this, we show that by considering the orientation-reversing quotient in a larger $\mathbb R^{2,5}$ embedding spacetime, we can derive a similar a $\ds\times\reals$ quotient spacetime. Using the known conformal relationship between Rindler wedges of $\ads\times\twosphere$ and static patches of $\ds\times\R$ embedded in such an $\mathbb R^{2,5}$, we argue that the structure of the conformal vacuum of Rindler $\noads\times\twosphere$ is reflected in the vacuum structure of a novel $\ds\times \reals$ whose asymptotic boundaries are glued together to another $\ds\times\reals$. We construct this novel $\ds\times\reals$ in section \ref{sec:nodstimesR}. Specifically, we argue that the conformal relationship relates the conformal vacuum supported on two sides of a single $\noads\times\twosphere$ Rindler wedge boundary to  the conformal vacuum supported on two equivalent static patches in the two glued $\ds\times\reals$ that make up $\nodstimesR$. 

In section \ref{sec:noads} we review the construction of $\noads$ as a quotient of two AdS bulks glued along their conformal boundary, discuss features of its topology and its holography. We then demonstrate how $\noads\times\twosphere$ can be constructed via embedding in a non-orientable $\mathbb R^{2,5}$ lightcone. In section \ref{sec:nodstimesR}, we show how a dS$_4$ product space is constructed via embedding into the same $\mathbb R^{2,5}$ lightcone, and discuss its topology. We then discuss the conformal relationship between its static patches and Rindler wedges of $\noadstwosphere$, and its implications for their conformal vacua. We conclude in section~\ref{sec:discussion} with a discussion about holography in the de Sitter space.

\section{$\noads$}
\label{sec:noads}

In this section we review the construction of non-orientable $\ads$ ($\noads$) from a quotient of two $\ads$'s glued together at their conformal boundaries. We discuss its bulk and boundary topologies, and discuss the implications for its holography.

\subsection{Construction}\label{sec:NOAdS Construction}

$\ads$ is commonly embedded in $\RR{2}{2}$ via a hyperboloid constraint of the form $-u^2-v^2+x^2+y^2=-\ell^2$. The metric on $\RR{2}{2}$ is $\dd s^2 = -\dd u^2-\dd v^2+\dd x^2+\dd y^2$, and the AdS metric is obtained as the induced metric on the embedded hyperboloid. The hyperboloid can be parameterized in global coordinates as
\begin{align}\label{eq:ads3_extended}
  u &= \ell \sec(\rho) \cos(t), \quad v = \ell \sec( \rho) \sin(t),\\
  x &= \ell \tan( \rho) \cos(\phi), \quad y = \ell \tan(\rho) \sin(\phi),
\end{align}
where $\rho \in[0,\pi/2)$, $\phi \in [0,2 \pi)$, and $t \in [0,2\pi)$ (or $ t\in \mathbb R$ for the universal cover of $\ads$). The induced AdS metric is

\begin{equation}
  \dd s^2 = \frac{\ell^2}{\cos^2 \rho} (-\dd t^2 + \dd  \rho^2 + \sin^2 \rho \dd \phi^2).
\end{equation}
This space has topology $\onesphere\times\text{D}_2$ (or $\R\times\text{D}_2$ for its universal cover).

\begin{figure}
    \centering
    \includegraphics[width=0.7\linewidth]{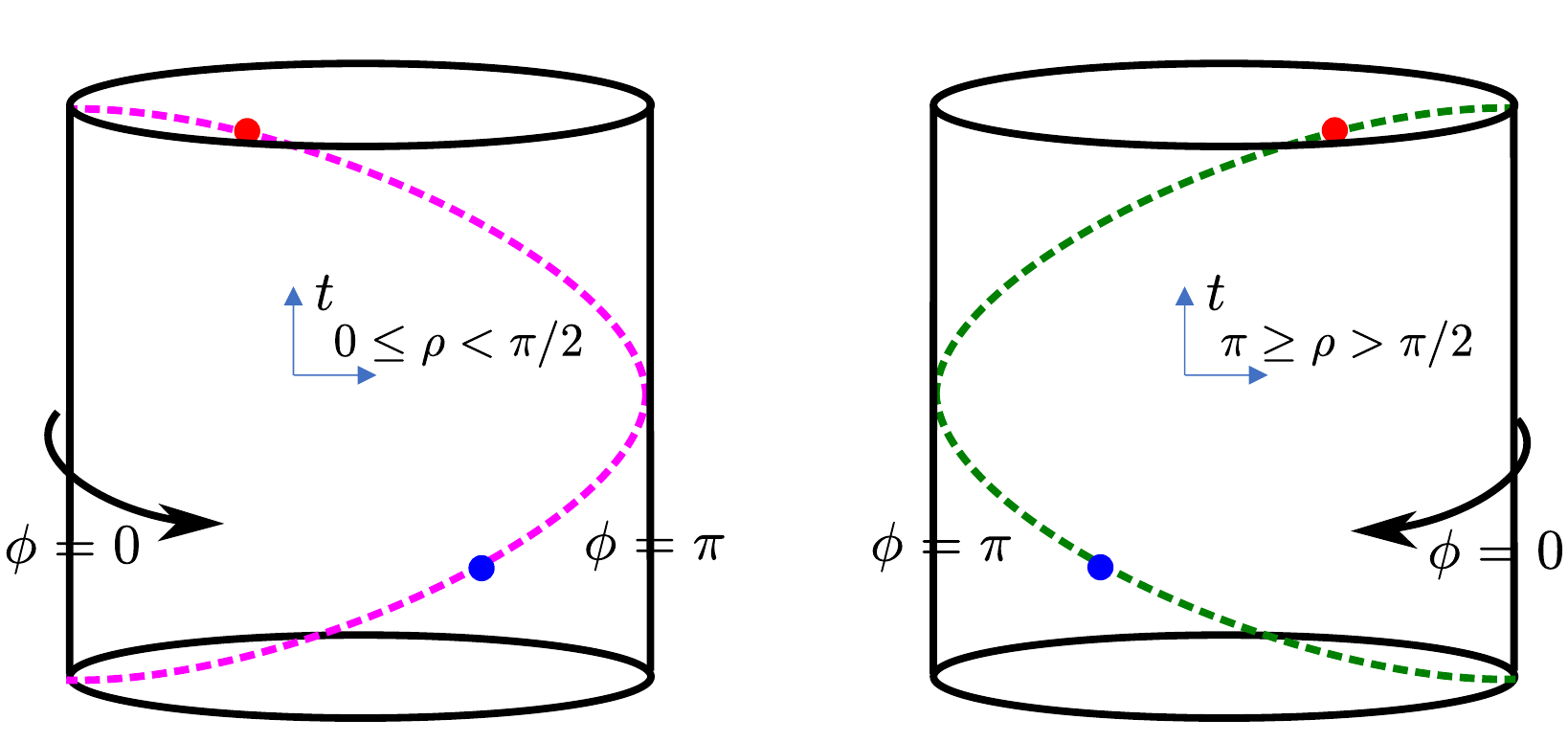}\\
    \includegraphics[width=0.9\linewidth]{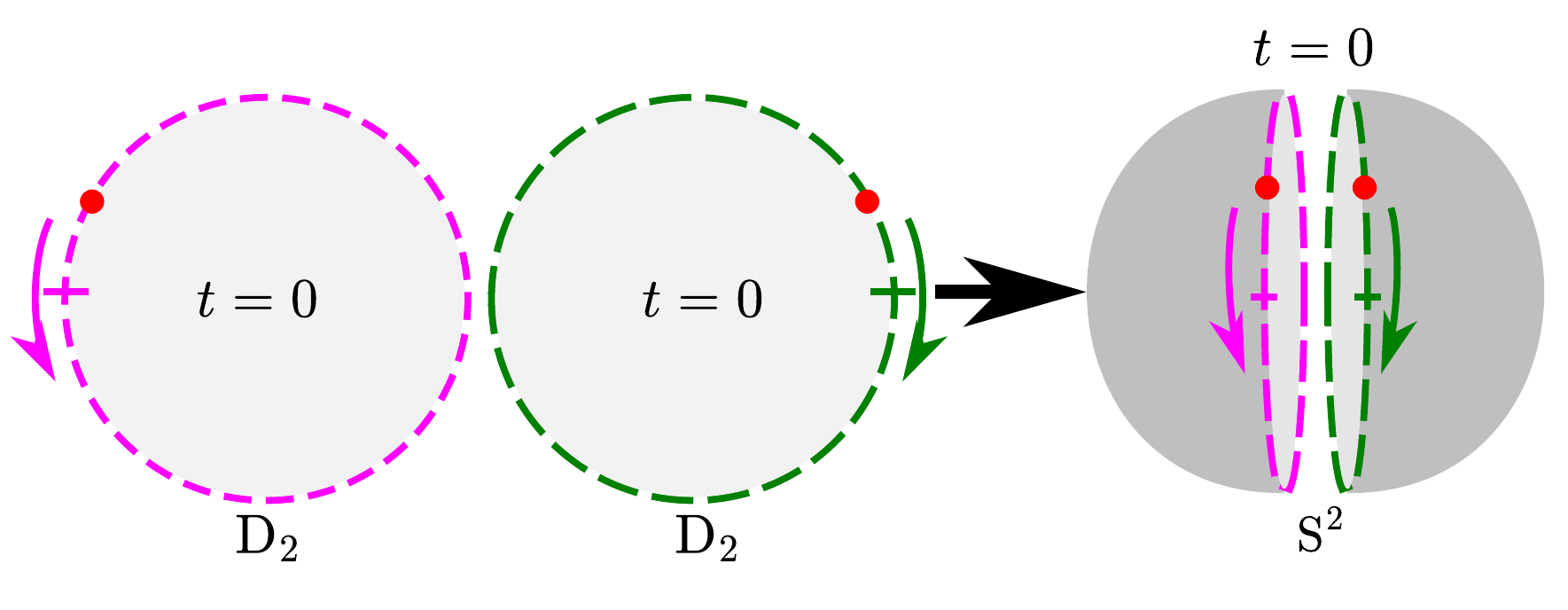}
    \caption{ a) An example of a chiral curve (pink) on the boundary of AdS which is a glued to a corresponding curve (green) on the extended AdS. The $(t,\rho,\phi)$ chart on the two spaces have opposite handedness, indicated by the apparently opposite chirality of the two boundary curves. Red and blue dots indicate specific points which glued to one another. b) The boundaries of the $t=0$ slice are glued together, forming an $\twosphere$ topology.}
    \label{fig:ads_gluing}
\end{figure}

The standard compactification of $\ads$ is obtained by including the point $\rho = \pi/2$. We further extend the space to include a second bulk region defined on $\pi/2<\rho\leq\pi$. This extension corresponds to gluing two $\ads$ spaces together along their conformal boundaries at $\rho = \pi/2$ to produce a boundary-less spacetime whose topology becomes $\onesphere\times\twosphere$ \cite{Arias:2025vjy}. As in regular $\ads$, the $\onesphere$ corresponds to the timelike direction, whereas the $\twosphere$ is obtained by gluing the two spacelike $\text{D}_2$s along their boundaries. 
Also note that since the original AdS is covered by $\rho \in [0,\pi/2)$ with the radially-outward direction along the $+\rho$-axis, and the extended region is covered by $\rho\in (\pi/2,\pi]$ with the radially-outward direction along the negative $\rho$-axis, the two AdS spaces are oppositely oriented with respect to one another. This is depicted in Fig~\ref{fig:ads_gluing}, where the extended AdS bulk is given a left-handed frame.

\begin{figure}[htbp]
    \centering
    \includegraphics[width=0.7\linewidth]{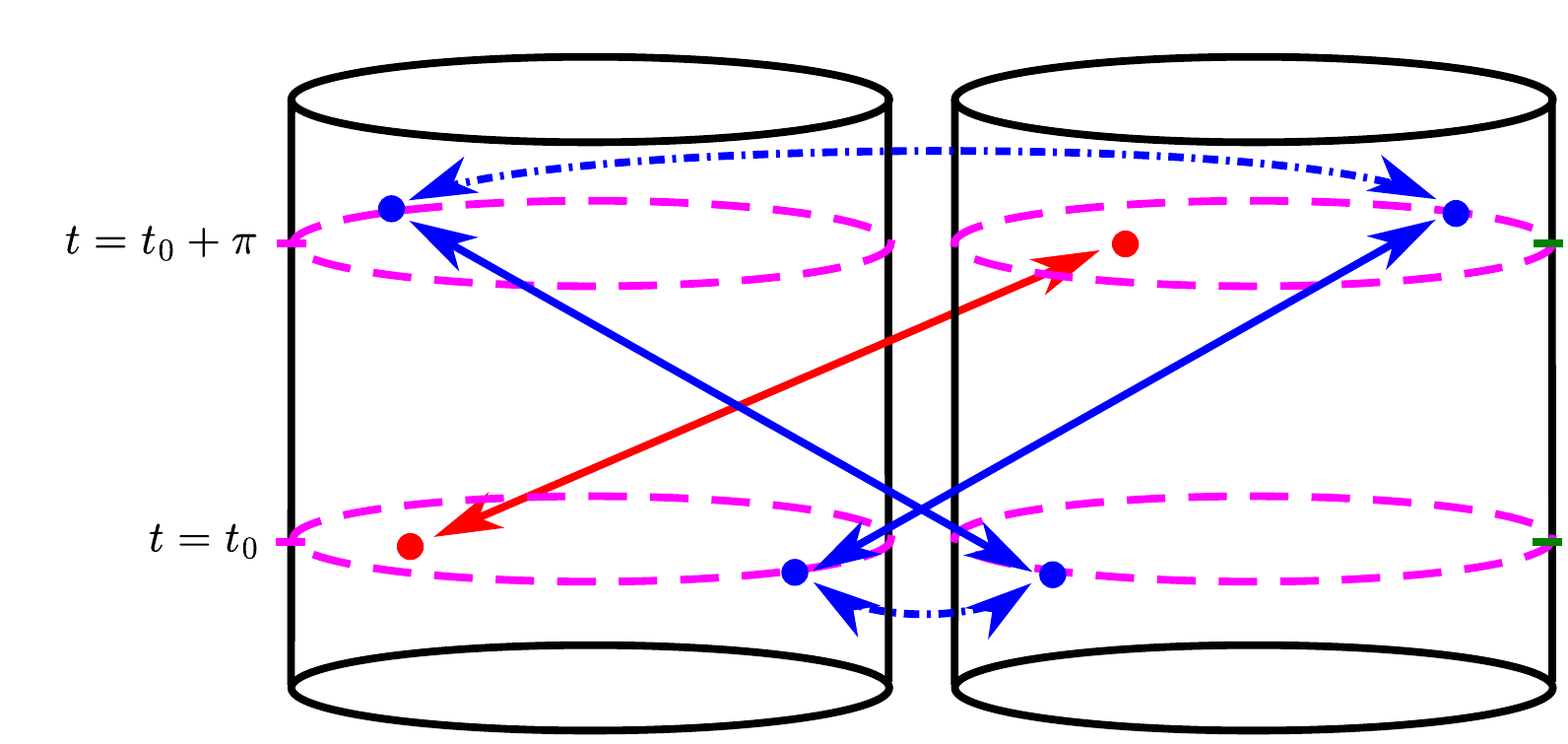}
    \includegraphics[width=0.7\linewidth]{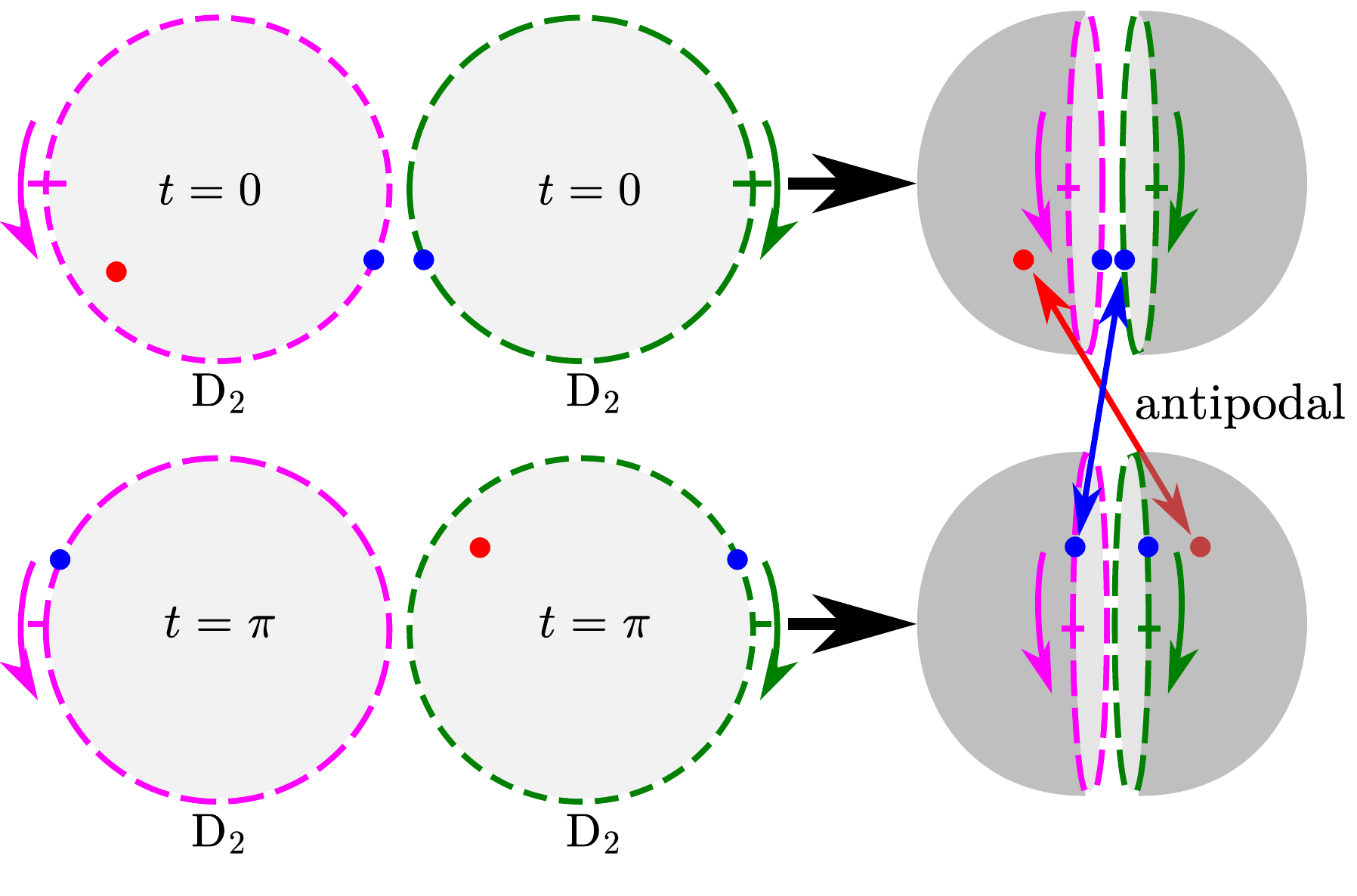}
    \caption{a) Points in the bulk (red) are identified by $\Sigma$, indicated by a solid line. Points on the boundary (blue) are identified both under $\Sigma$, and via the gluing rule, indicated by a double line.  b)Sigma produces a twisted product topology: it's action is $\pi$ shift in along the $\onesphere$ factor, combined with antipodal identification on the $\twosphere$. }
    \label{fig:sigma_quotient}
  \end{figure}

Quotienting this boundaryless spacetime by a free $\mathbb Z_2$ action defined by the map $\Sigma(t, \rho, \phi) = (t+\pi, \pi -\rho, \phi + \pi)$ produces a non-orientable boundaryless $\ads$ spacetime~\cite{Pathak24}, referred to as $\noads$. The resulting space has a twisted product topology $\onesphere\tildetimes\twosphere$, where points at time $t=t_0$ are identified with their antipode (on the compact spacelike $\twosphere$) at time $t= t_0+\pi$. To see that the the spatial identification is antipodal, note that the $\rho$ coordinate plays the role of the polar angle, and $\phi$ that of the azimuthal angle, so that the combined action of $\rho\rightarrow \pi-\rho$ and $\phi\rightarrow \phi+\pi$ corresponds to a combined reflection across the polar equator and a azimuthal shift by $\pi$.  
See Fig.~\ref{fig:sigma_quotient} for a visual representation of the gluing rule, quotient, and resulting twisted $\onesphere\tildetimes\twosphere$ topology. 

\begin{figure}[htbp]
    \centering
    \includegraphics[width=1.\linewidth]{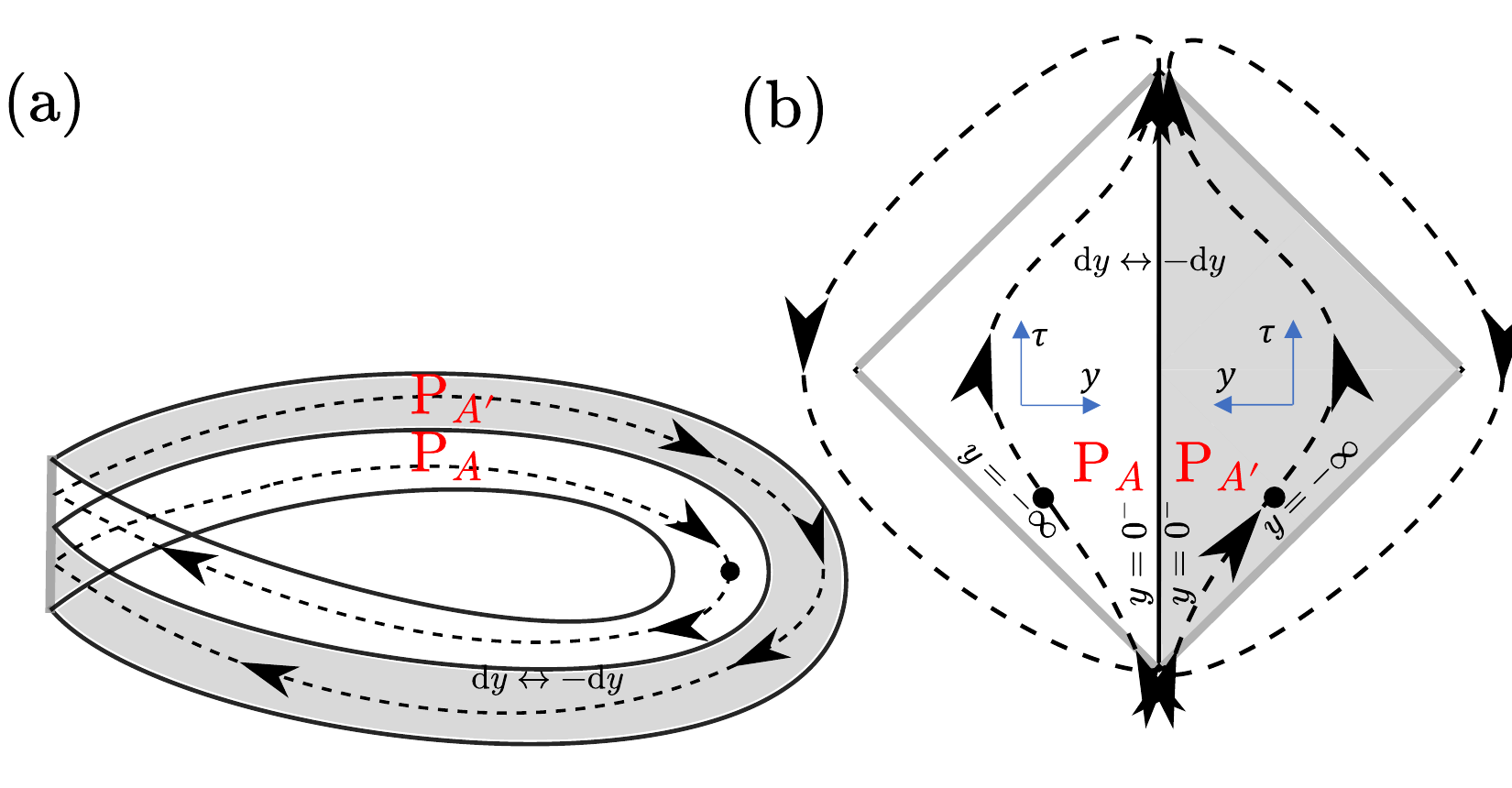}
    \caption{$\noads$'s topology consists of north and south hemispheres of $\twosphere$s glued at their equator, $\pi$ apart in time. This is equivalent to gluing the two Poincar\'e patches of $\ads$ together at their boundary, creating an ``internal'' boundary, such that the bulk twists by $\pi$ at the Poincar\'e horizons. As a result, (a) they contain a M\"obius strip topology where each oppositely-oriented Poincar\'e patch corresponds to one side of its centerline, where bulk timelike loops can only return to their initial point after a global time displacement of $\pi$ by crossing the centerline with opposite orientation. Otherwise they return to themselves with a displacement of $2 \pi$ (see dashed line) and never cross the centerline. (b) Such trajectories that return to themselves in $2 \pi$ traverse a bulk that is indistinguishably orientable to $\ads$ and so its Penrose diagram looks the same if you ignore its different gluing on its boundary.}
    \label{fig:noads}
  \end{figure}

  Note that despite being a ``boundaryless'' spacetime, it is still possible to identify conformal ``internal'' boundaries of $\noads$: in the coordinate system given earlier, they are located at $\rho = \pi/2$ and have topology $\onesphere \tildetimes \onesphere$. This internal boundary possesses a twisted torus topology, which is homeomorphic to a regular $\onesphere \times \onesphere$ torus. The remarkable property of this torus is that both $\onesphere$s are null, which means that this is a non-degenerate null torus (i.e.~it has a null fundamental lattice).

  We point out that since the boundary submanifold is homeomorphic to an ordinary torus, it is itself orientable. Likewise, the bulk spacetime, i.e. $\noads$ minus the points lying on the $\rho=\pi/2$ submanifold, also constitute an orientable manifold. It is only the combined compact manifold which is non-orientable. This is analogous to the statement that excising the center-line of a M\"obius strip yields two orientable strips.

  A M\"obius strip can be constructed via overlapping patches with transition functions, or alternatively as two non-overlapping patches with gluing rules (see Fig.~\ref{fig:noads}). The latter approach is natural for $\noads$, since $\Sigma$ exchanges Poincar\'e patches in one $\ads$ bulk with the complementary Poincar\'e patch in the other $\ads$.

  In regular $\ads$ the two Poincar\'e patches,
  \begin{align}
    \dd s^2 =
    \begin{cases}
      y^{-2}(-\dd \tau^2 + \dd x^2 + \ell^2 \dd y^2) &\text{ for } y > 0,\\
      y^{-2}(-\dd \tau^2 + \dd x^2 + \ell^2 \dd y^2) &\text{ for } y < 0
      \end{cases}
  \end{align}
where $\tau$, $x \in \reals$ and $(\dd \tau, \dd x, \dd y) \rightarrow (\dd \tau, \dd x, \dd y)$ at $y \rightarrow \pm \infty$ and $t \rightarrow \pm \infty$.

On the other hand, for $\noads$, if P$_A$ and P$_{A'}$ are Poincar\'e patches in opposite $\ads$ bulks which share a boundary, and P$_B$, P$_{B'}$ are their respective complements, then $\Sigma$ identifies points in P$_A$ with points in P$_{B'}$, etc. The effect on the two Poincar\'e patches turns out to flip the tangent space (i.e.~volume element) with the gluing rule $(\dd \tau, \dd x, \dd y)\rightarrow (\dd \tau, \dd x, -\dd y)$ on their conformal boundary (see Appendix~\ref{app:Kleinextension}).

As a result, the full $\noads$ can be succinctly captured by a fundamental domain consisting of the two Poincar\'e patches (e.g.~P$_A$ and P$_B$) with their boundaries glued together with a twist ($\dd y \rightarrow -\dd y$):
  \begin{align}
    \dd s^2 =
    \begin{cases}
      y^{-2}(-\dd \tau^2 + \dd x^2 + \ell^2 \dd y^2) &\text{ for } y > 0,\\
      y^{-2}(-\dd \tau^2 + \dd x^2 + \ell^2 \dd y^2) &\text{ for } y < 0
      \end{cases}
  \label{eq:NO-AdS3-patches}
\end{align}
where $\tau$, $x \in \reals$ and $(\dd \tau, \dd x, \dd y) \rightarrow (\dd \tau, \dd x, \dd y)$ at $y \rightarrow \pm \infty$ and $t \rightarrow \pm \infty$ (i.e.~the same gluing rule as for $\ads$), and most importantly, at their conformal boundary $(\dd \tau, \dd x, \dd y) \rightarrow (\dd \tau, \dd x, -\dd y)$, which becomes an ``internal'' boundary of a boundaryless spacetime~\footnote{Note that is possible to ``push'' this twist to the bulk by diffeomorphically transforming one of the Poincar\'e patches and exchanging their $\tau$ and $x$ coordinates (i.e.~the diffeomorphism induces an additional twist at both the horizons and the boundary gluing rules)~\cite{Pathak24}.}. As a result, a timelike observer that traverses a loop from one Poincar\'e patch to another, and then travels through the conformal ``internal'' boundary back to its initial point, returns to itself with opposite orientation in its tangent space~\cite{Pathak24} (see Fig.~\ref{fig:noads}a).

We are free to choose as our fundamental domain of $\noads$, Poincar\'e patch P$_A$ and P$_B$ as in Eq.~\ref{eq:NO-AdS3-patches}, or Poincar\'e patches P$_A$ and P$_{A'}$. It turns out, much like in the similar $\onesphere\times\RP{2}$ spacetime, that both fundamental domains appear identical due to the lack of an external boundary. A Penrose diagram of this spacetime with Poincar\'e patches P$_A$ and P$_{A'}$ forming the fundamental domain is shown in Fig.~\ref{fig:noads}b, though this is not strictly a compactification to the Einstein static universe due to the ``internal'' boundary.

\begin{figure}[htbp]
    \centering
    \includegraphics[width=1.\linewidth]{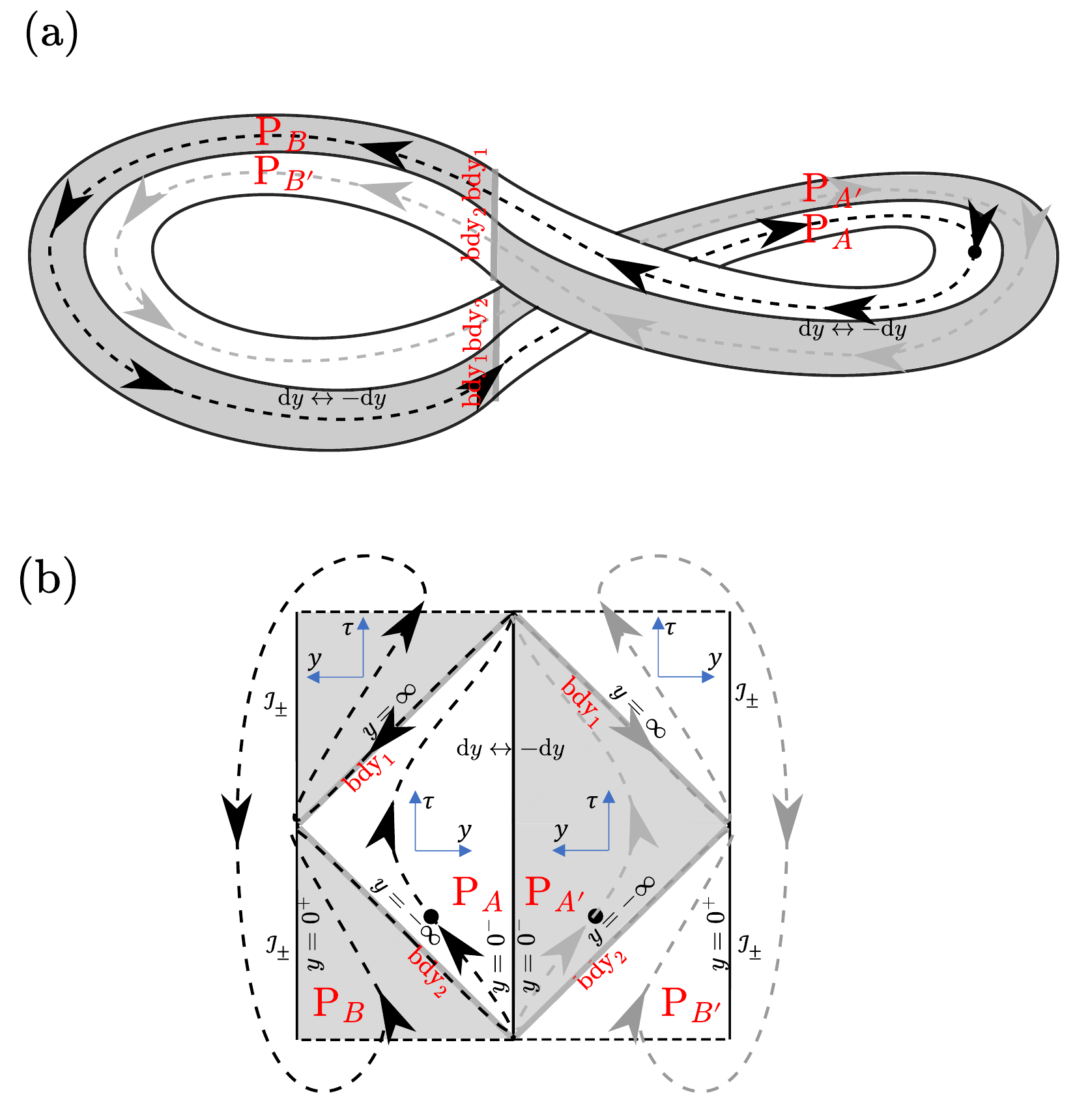}
    \caption{(a) $\noads$'s double cover undoes the $\Sigma$ quotient and corresponds to two $\ads$s glued together at their boundary producing an $\onesphere \times \twosphere$ topology. The conformal boundaries of each Poincar\'e patch in one $\ads$ are shared with the same Poincar\'e patch in the other $\ads$. (b) Drawn are the two timelike trajectories in the Penrose diagram that equivalent to the one from before in Fig.~\ref{fig:noads}.}
    \label{fig:noadsdoublecover}
  \end{figure}

The universal cover of $\noads$ can be obtained by unrolling $t$ (see Fig.~\ref{fig:noadsdoublecover}). This necessarily undoes the $\Sigma\equiv \mathbb Z_2$ orientifold quotient and produces an $\reals\times\twosphere$ topology. Note that this universal cover is equivalent to two glued universal covers of $\ads$. This space has the topology of two $\reals \times \text{D}_2$ cylinders glued along their $\R\times\onesphere$ boundaries. As a result, the remaining compact dimension on the ``internal'' boundary of the universal cover is spacelike.

\begin{figure}
    \centering
    \includegraphics[width=1.0\linewidth]{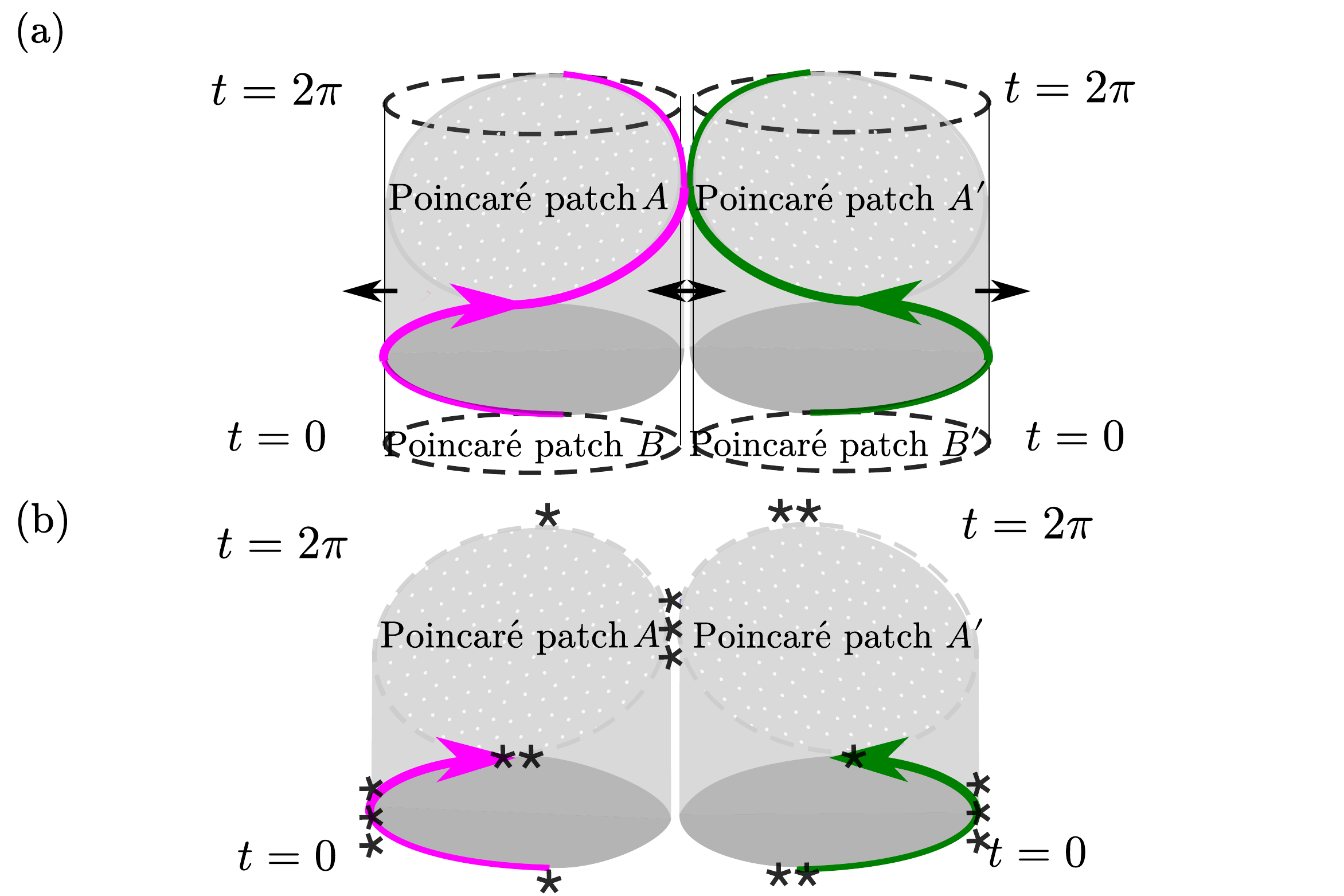}
    \caption{A sketch of (a) two $\noads$ cylinder's glued together at their boundary, aligning along Poincar\'e patches, producing a topology of $\onesphere \times \twosphere$. As a result of the gluing, chiral and anti-chiral components (i.e.~green and pink curves) of one Poincar\'e patch become equal to anti-chiral and chiral components in the other Poincar\'e patch.
      The $\Sigma$ orientation-breaking quotient identifies the two $\ads$ with exchanged Poincar\'e patches, (b) producing a fundamental domain of $\noads$ consisting of two Poincar\'e patches glued together at their conformal boundary ($t\rightarrow t+\pi$ and $\phi\rightarrow \phi+\pi$). Trajectories that leave the top (bottom) of A or A$'$ enter the bottom (top) of A$'$ or A, respectively, rotated by $\pi$ with respect to each other (denoted by the `$\star$', `$\star\star$' and `$\star\star\star$'). The resultant topology is non-orientable, boundaryless and compact $\onesphere \tildetimes \twosphere$. This means that its bulk by itself is indistinguishable from the bulk of $\ads$, except that it wraps around both sides of its boundary, which has the topology of a torus with a null periodic lattice. However, as an embedded manifold, this null-periodic torus is one-sided, which means that chiral (green) trajectories on one side of the boundary end on anti-chiral (pink) trajectories on the other side of the boundary after a $\pi$ loop, and vice-versa, and are thus $2 \pi$ periodic.}
    \label{fig:Rindlerboundaries}
  \end{figure}

  The quotient $\noads$ spacetime still contains chiral and anti-chiral directions, however, on the boundary they have half the domain of the pre-quotiented spacetime. This allows them to be globally independent, making the internal boundary a Lorentzian torus with a null-periodic lattice instead of a lattice that is periodic along timelike and spacelike directions (i.e.~the ``internal'' boundary topology is null $\onesphere\times$ null $\onesphere$, both with period $\pi$). However, as an embedded manifold, this null-periodic torus is a one-sided torus with $2 \pi$ period. This means that while it has a trivial holonomy in its tangent space, and is thus orientable, it has a non-trivial $\mathbb Z_2$ holonomy in its normal space, which means that the embedding wraps around both sides of it. We will explore the holographic consequences of this factorization in the next section.

\subsection{Holography}
\label{sec:noads-holography}

Before turning to the holography of $\noads$, we first discuss the holography of two boundary-glued copies of $\ads$, so that we may contrast their features. We first consider two universal covers of $\ads$, say $\widetilde{\mathcal{M}}^+$ and $\widetilde{\mathcal{M}}^-$ glued along their asymptotic boundaries, i.e. $\partial\widetilde{\mathcal{M}}^+=\partial\widetilde{\mathcal{M}}^-\equiv\widetilde{\mathcal{B}}$, but which are otherwise non-interacting. Here tilde denotes the universal cover. We suppose that the shared boundary maintains the usual Dirichlet (reflecting) boundary conditions. We require that any metric fluctuations between the independent bulk theories approach the same (fixed) asymptotic metric, but no other dynamical data is transmitted.  The shared boundary therefore supports the asymptotic degrees of freedom of both bulk theories. The gluing surface $\widetilde{\mathcal{B}}$ can thus be regarded as supporting two identical non-interacting CFTs $H_{CFT}\otimes H_{CFT}$. The ground state is simply a product of ordinary $\ads$ ground states: $\Psi_{vac} = \psi_{vac}\otimes\psi_{vac}$.  
    
In the case where two single-cover $\ads$ spaces are glued (as opposed to their universal covers), e.g. so that $\partial\overline{\mathcal M}^+ = \partial\overline {\mathcal M}^- = \overline{\mathcal{B}}$, the existence of closed timelike curves alters the structure of the theory. Specifically, the conformal boundaries are glued on a Lorentzian torus, $\overline{\mathcal{B}}$, with a time- and space-periodic lattice with individual periods $2\pi$. We assume that the asymptotic dynamics remain describable by an appropriately time-periodic 2D CFT, analogous to the wedge $\text{CFT}_2$ duals discussed in \cite{Melton:2025ecj}. Due to the non-interaction of the theories, we still regard the gluing surface $\overline{\mathcal{B}}$ as supporting two such CFTs. The vacuum state maintains factorization into a product of two CFT vacua. Going forward, we use $H^{\overline{\mathcal{B}}}$ to refer to the putative time-periodic CFT$_2$ of the single-cover theory. When we wish to specifically refer to this putative CFT$_2$ on a particular $\pm$ orientation (or global ``side'') of its support, we use $H^{\overline{\mathcal{B}}}_\pm$. 

This factorization is maintained when restricting to subregions of the boundary. For example, if $\Omega$ denotes a causal diamond on the shared boundary $\overline{\mathcal{B}}$ with two bulk domains of dependence $\omega^+$ and $\omega^-$ on opposite sides, then the state $\rho_\Omega\in H^{\overline{\mathcal{B}}}_{\Omega^+}\otimes H^{\overline{\mathcal{B}}}_{\Omega^-}$ must factorize. In particular, it is dual to a bulk state supported on $\omega^+\,\cup\,\omega^-$: 
\begin{equation}\label{eq:pre-quotient}
\rm{Pre-quotient:}\qquad \rho_{\Omega} \in H^{\overline{\mathcal{B}}}_{\Omega^+}\otimes H^{\overline{\mathcal{B}}}_{\Omega^-}\quad\xleftrightarrow[]{CFT /AdS} \quad \rho_{\omega^+\,\cup\,\omega^-} = \rho_{\omega^+}\otimes\rho_{\omega^-}
\end{equation}
In the above expressions, $H^{\overline{\mathcal{B}}}_{\Omega}$ refers to the Hilbert space obtained by restricting the original Hilbert space factors $H^{\overline{\mathcal{B}}}$ to the subregion $\Omega$, and its two sides $\Omega^\pm$.

We now turn to the effect of the $\mathbb Z_2$ quotient, which identifies points $(t,\rho,\phi)$ in $\overline{\mathcal M}^+$ with the points $(t+\pi,\pi-\rho,\phi+\pi)$ in $\overline{\mathcal M}^-$. This has the effect of modifying the global topology to that depicted in Fig~\ref{fig:noads}, where the $\ads$ bulks, formerly connected only superficially through their boundary gluing, are now connected through the bulk directly. The boundary topology is also modified to be that of a Lorentzian torus with null fundamental lattice, which we denote by $\mathcal B$. We regard this boundary torus $\mathcal B$ as still supporting two non-interacting boundary theories, though they will inherit a permutation-twisted structure discussed shortly.

Consider fields $\mathcal O_\pm$ defined on opposite sides of $\overline{\mathcal{B}}$, before the quotient. The $\Sigma$ action can be represented by a permutation $P$ holonomy on this original double cover space, $\hat x \in \overline{\mathcal{B}}$, since the normal-vector flips:
\begin{equation}
P \mathcal O_\pm(\hat x) \equiv \mathcal O_\pm(\Sigma \hat x) = \mathcal O_\mp(\hat x).
\end{equation}

Now restrict the fields $\mathcal O_\pm$ to the torus $\mathcal{B}$ after the quotient with null-periodic lattice vectors $v_\pm=x\pm t$ and periods $\pi$. Since this torus is produced by the $\Sigma$ quotient, it follows that points go to themselves under $\Sigma$. In fact, action by $\Sigma$ is equivalent to translation by $v_\pm$. Nevertheless, the effect on the $\mathbb Z_2$-fiber of the normal space can still be kept by staying in this permutation-twisted sector~\cite{Borisov98}.
\begin{equation}
  \left(\begin{matrix} \mathcal O_+\\ \mathcal O_- \end{matrix} \right) (x + v_\pm) = P \left( \begin{matrix} \mathcal O_+\\ \mathcal O_- \end{matrix} \right) (x),
\end{equation}
where $x \in \mathcal{B}$ and $P = \left( \begin{matrix} 0& 1\\ 1& 0\end{matrix}\right)$ is the permutation matrix.

In this permutation-twisted sector, the CFT$_2$ Hilbert spaces on the two sides of the null-periodic torus are not individually periodic under the null lattice vectors. Rather, the two-copy system is \emph{globally} defined on the null-periodic torus with a permutation transition function.

$\overline{\mathcal B}$ is double cover of $\mathcal B$. This means that $P$-twisting around both null cycles of $\mathcal B$ reconstructs the original time-space-periodic  $\overline{\mathcal{B}}$ (see Fig.~\ref{fig:Rindlerboundaries}), which implies that the permutation-twisted Hilbert space, $\left(H^{\mathcal{B}}_+ \otimes H^{\mathcal{B}}_- \right)_{P-\text{twisted}}$, is periodic under a lattice with the same primitive lattice vectors as supports $H^{\overline{\mathcal{B}}}_\pm$. In this sense, the $P$-twisted construction is similar to one of the two original CFT$_2$ Hilbert spaces on the time-space-periodic torus $\overline{\mathcal{B}}$. This aligns with the expectation that the $\Sigma=\mathbb Z_2$ quotient should take the two CFT$_2$s defined on the original two-sided time-space-periodic torus (with Hilbert spaces $H^{\overline{\mathcal{B}}^\pm}_{\text{CFT}}$) and produce one new CFT$_2$ defined on the same manifold.

Notice that this permutation-twisted sector is very similar to what occurs when a CFT$_2$ is defined on a cross-cap and its flip is in the spatial dimensions. Therein, the Virasoro primaries are identified $L_m \leftrightarrow (-1)^m \bar L_{m}$. Here, the $P$-twisted construction instead identifies between two CFT$_2$ chiral (and antichiral) Virasoro operators, $L^{\mathcal{B}_+}_{m} \leftrightarrow L^{\mathcal{B}_-}_m$ (and c.c.), instead of chiral and anti-chiral operators in a single CFT$_2$.

An immediate consequence of the $P$-twisted two-copy sector is that CFT$_2$ states cannot factorize into $H^\mathcal{B}_{+} \otimes H^\mathcal{B}_{-}$. If the AdS/CFT correspondence holds for wedge CFT$_2$s, this forbidden factorization should be inherited by the bulk. Indeed, this turns out to occur since the bulk on opposite sides of the boundary $\mathcal B$ becomes causally connected after the quotient. Generically, this is because field data on P$_{A'}$ in Fig.~\ref{fig:noads} is related to field data in P$_A$ through bulk time evolution. We thus expect that a single state of the boundary theory, say $\rho_{CFT}\in \left(H^\mathcal{B}_{+} \otimes H^\mathcal{B}_{-} \right)_{P-\text{twisted}}$ is dual to an (entangled) bulk state supported on slices through both P$_A$ and P$_{A'}$ 

As a specific example, consider, as in Eq.~\ref{eq:pre-quotient}, a boundary subregion $\Omega$ on $\mathcal B$ after the quotient is performed.  Its bulk domain of dependence still consists of two Rindler wedges $\omega^+$ and $\omega^-$ on opposite sides of $\mathcal B$. However, since the regions $\omega^+$ and $\omega^-$ intersect with each others' causal future/past, the state supported on $\omega^+\,\cup\,\omega^-$ is non-trivially entangled. 
\begin{equation}\label{eq:post-quotient}
    \rm{Post-quotient:}\qquad\rho_\Omega\in\left(H^\mathcal{B}_{\Omega^+} \otimes H^\mathcal{B}_{\Omega^-} \right)_{P-\text{twisted}}\xleftrightarrow[]{ CFT/AdS }  \rho_{\omega^+\,\cup\,\omega^-}\neq \rho_{\omega^+}\otimes\rho_{\omega^-}
\end{equation}

In a standard Lorentzian torus the chiral and anti-chiral components are locally decoupled, leading to independent chiral and anti-chiral Virasoro algebras. However, globally, since the two null directions are not independent in $\Sigma$ (i.e.~$-\pi\le v_+ < \pi$ and $-|u|+\pi \le v_- < |u|-\pi$), global boundary conditions couple these two components together. Instead, $v_+\pm v_-$ are independent, which forces energy and spatial momentum to be quantized,
\begin{equation}
  E = P_R + P_L = \frac{2 \pi m}{L}, \quad  P = P_R - P_L = \frac{2 \pi n}{L},
\end{equation}
for $m$, $n \in \mathbb Z$. As a result, quantum states and partition functions do not factorize into global chiral and anti-chiral terms since
\begin{equation}
  Z(\tau, \bar \tau) = \tr_{\mathcal H} (q^{L_0 - \frac{c}{24}} \bar q^{\bar L_0 - \frac{c}{24}}), 
\end{equation}
where $q = e^{2 \pi i \tau}$ and $\tau = \frac{\alpha_2}{\alpha_1}$ is the modular parameter for a torus with a lattice generated by $(\alpha_1, \alpha_2)$ along $\vec u$ and $\vec v$, respectively.

On the other hand, in a torus with a null-periodic lattice, boundary conditions are decoupled (i.e.~$-\pi\le v_+<\pi$ and $-\pi\le v_-<\pi$) and so their spectra are completely independent,
\begin{equation}
  P_L = \frac{2 \pi n}{L}, \quad P_R = \frac{2 \pi m}{L},
\end{equation}
where $m$, $n \in \mathbb Z$. This means that quantum states and the partition function factorize,
\begin{equation}\label{eq:chiral_factor}
  Z(\tau, \bar \tau) = Z_{\text{chiral}}(\tau) Z_{\text{anti-chiral}}(\bar \tau).
\end{equation}

This simplification into two globally independent (or ``non-interacting'') chiral and anti-chiral factors is not possible in higher dimensions and signifies that $\noads$ cannot be generalized to higher dimensions. This also can be seen from a topological argument: while it is always possible to glue two AdS$_{d+1}$ together at their boundary to produce timelike S$^1 \times$ S$^{d}$, and subsequently quotient by the $(d+1)$-dimensional generalization of $\Sigma$ to produce S$^1 \tildetimes$ S$^d$, it is only for even $d$ that the resultant compact spacetime is non-orientable (since antipodal identifications are reflections in even $d$ and rotations in odd $d$). Restricting to even $d$, it follows that the interior boundary of these quotiented spacetimes can be identified as an orientable $\onesphere \tildetimes$ S$^{d-1}$, which supports a CFT$_d$. While this interior boundary is homeomorphic to $\onesphere \times$ S$^{d-1}$ where the $\onesphere$ is null, only for $d=2$ is the S$^{d-1}$ also null instead of spacelike. In any case, a CFT$_d$ does not factor into two independent algebras for $d>2$ let alone two independent global groups.

A note of caution: while it is true that in terms of its internal boundary the single cover of $\noads$ has a torus boundary with null-periodic lattice, its universal cover does not transform this to a cylinder with a null compact direction. Specifically, the universal cover of $\noads$'s $\onesphere \tildetimes \twosphere$ topology involves unrolling the timelike $\onesphere$ to $\reals$. The internal boundary does not ``respect'' this since it is not a true external boundary, and so can actually be found to become timelike $\reals$ times spacelike $\onesphere$: the usual timelike cylinder with spacelike compact direction. This is because the universal cover is equivalent to just gluing together the universal covers of two $\ads$ at their boundaries.

To study the $\ds$ fiber bundle in the next section, we will use a conformal map between $\ds\times\R$ static patches and $\noads\times\twosphere$ Rindler wedges. We show in Appendix~\ref{app:noadstwosphere} that the $\twosphere$ fiber does not affect the asymptotic group structure compared to $\noads$, because the Brown-Henneaux conditions are satisfied for this space. Moreover, we do not expect the entanglement between the boundary-glued $\noads\times\twosphere$ Rindler wedges $\omega^+$ and $\omega^-$ in the state $\rho_{\omega^+\,\cup\,\omega^-}$ to differ from those of $\noads$, though the structure within each $\omega^+$ or $\omega_-$ will generically differ due to the $\twosphere$ factor.

\section{$\nodstimesR$}
\label{sec:nodstimesR}

 In this section, we construct an asymptotically de Sitter spacetime with future and past asymptotic boundaries that are glued together. We discuss its features and its relationship to $\noadstwosphere$. In Sec.~\ref{subsec:nodstimesRconstruction} we construct the space using a modified $\mathbb R^{2,5}$ lightcone embedding. We describe this space as two de Sitter global spacetime which are glued together at their two similar future and past boundaries. We call this spacetime $\nodstimesR$, where the ``BG'' stands for ``boundary-glued''. We then discuss the consequences of the geometry for the static patch vacuum. In Sec.~\ref{sec:dscftcorrespondence} we discuss the relationship between the static patch vacuum of $\nodstimesR$ and the Rindler vacuum of $\noadstwosphere$, highlighting the entanglement structure outlined in Sec.~\ref{sec:noads-holography}, and comment on a potential holographic description of $\nodstimesR$.
 
\subsection{$\nodstimesR$ Construction}
\label{subsec:nodstimesRconstruction}

$\ads$ can be described as a hyperbolic restriction on $\RR{2}{2}$: $-u^2 - v^2 + x^2 + y^2 = -\ell^2$. It follows that
$\ads\times\twosphere$ can be described by an analogous restriction to $\RR{2}{5}\simeq \RR{2}{2}\times\R^3$:
\begin{align}
  \label{eq:noadstwosphererestriction1}
    \underbrace{-u^2 -v^2 + x^2 + y^2}_{=-\ell^2} + \underbrace{a^2+b^2 +c^2}_{=+\ell^2}&= 0,
\end{align}

Considering a ``conformal extension'' of $\RR{2}{2}$ allows for $\noads$ to be derived through a hyperbolic restriction (see Appendix~\ref{app:Kleinextension}). $\noadstwosphere$ can be obtained via the same restriction after conformally extending the 
the $\RR{2}{2}$ factor of $\RR{2}{5}$. 

Rather than embedding $\ads\times\twosphere$ via the restrictions given by Eq.~\ref{eq:noadstwosphererestriction1}, we instead impose $\ds\times\R$ restrictions via the following constraint equation: 
\begin{align}
  \label{eq:dstimesRconstraint1}
    \underbrace{-u^2+ x^2+a^2+b^2+c^2 }_{=+\ell^2}  \underbrace{-v^2+y^2}_{=-\ell^2}&= 0.
\end{align}

Such an operation extends $\ds\times\R$ smoothly across its conformal boundaries gluing it future conformal boundary to a second future conformal boundary etc.

This can be seen by defining a global set of coordinates of $\ds$ in $\RR{1}{4}$ coordinatized by $(u,(\vec X,x))$ by 
\begin{align}
  \label{eq:dSglobal1}
  u_\pm &= \ell \sinh(t/\ell),\\
  \label{eq:dSglobal2}
  X_i &= \ell \cosh(t/\ell) \sin(\theta) \cos(\varphi_i) \prod_{j=1}^{i-1} \sin(\varphi_j), \text{ where } i=1,2,3 \text{ and}\\
  \label{eq:dSglobal3}
  x_\pm &= \ell \cosh(t/\ell) \cos(\theta),
\end{align}
and a hyperbola on the remaining $\RR{1}{1}$ factor of the $\RR{2}{5}$, as dictated by our restriction:
\begin{equation}
  \label{eq:dSglobal4}
  v_\pm = \ell \cosh \chi, \quad y_\pm = \ell \sinh \chi,
\end{equation}
such that 
\begin{equation}
  \dd s^2 = - \dd t^2 + \ell^2 \cosh^2(t/\ell) \dd \Omega_3^2 + \ell^2 \dd \chi^2,
  \label{eq:dStimesRglobalmetric}
\end{equation}
where the $\pm$ subscript denotes two branches which are related to each other via conformal extension from $u_+$, $v_+$, $x_+$, and $y_+$ at $\pm \infty$ to $u_-$, $v_-$, $x_-$, and $y_-$ at $\pm\infty$, respectively. This means that it extends each Poincar\'e patch of an embedded $\ds\times\R$ to another one which can be ``tracked'' in the above set of equations by taking the the second branch of the $\pm$.  

\begin{figure}
    \centering
    \includegraphics[width=1.\linewidth]{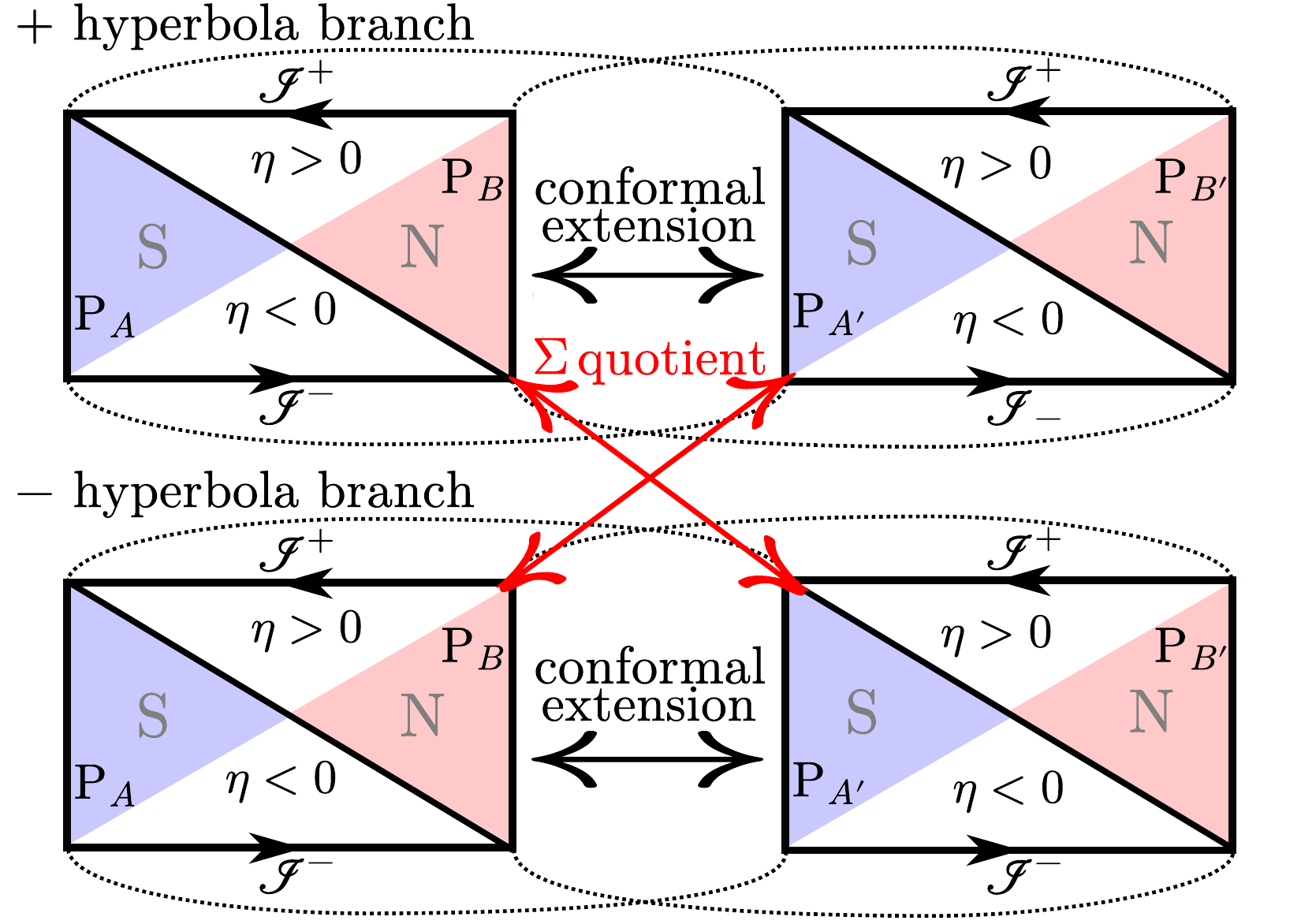}
    \caption{Penrose diagram of the two hyperbola branches of $\ds\times\mathbb H$ that are glued together by the conformal extension of the embedding space with two more $\ds\times\mathbb H$. The $\Sigma$ quotient then identifies $\pm$ hyperbola branches in the original $\ds\times \mathbb H^\pm$ antipodally with the $\mp$ hyperbola branches in the extended space, producing a quotient spacetime consisting of two $\ds\times\reals$ glued together at their similar past and future conformal boundaries.}
    \label{fig:dstimesRgluing}
\end{figure}

The conformal extension glues each $\ds\times \reals$ to another along the asymptotes of both the $\ds$ and $\reals$ factors.  Specifically, it can be seen from above that the $\ds\times \reals$ are glued together at $t,\, \chi \rightarrow \infty$ as well as at $t,\,\chi \rightarrow -\infty$, corresponding to a subset of both conformal boundaries. Notably, this means that, much like the case for $\ads$'s Poincar\'e patches, the $\ds$ Poincar\'e patches are also glued together at their conformal boundaries by the conformal extension, except that the asymptotic gluing surface is codimension-2 rather than codimension-1, since $\chi\rightarrow
\pm\infty$ in addition to $t\rightarrow\pm\infty$ (see Fig.~\ref{fig:dstimesRgluing})

\begin{figure}
  \centering
  \includegraphics[width=1.0\linewidth]{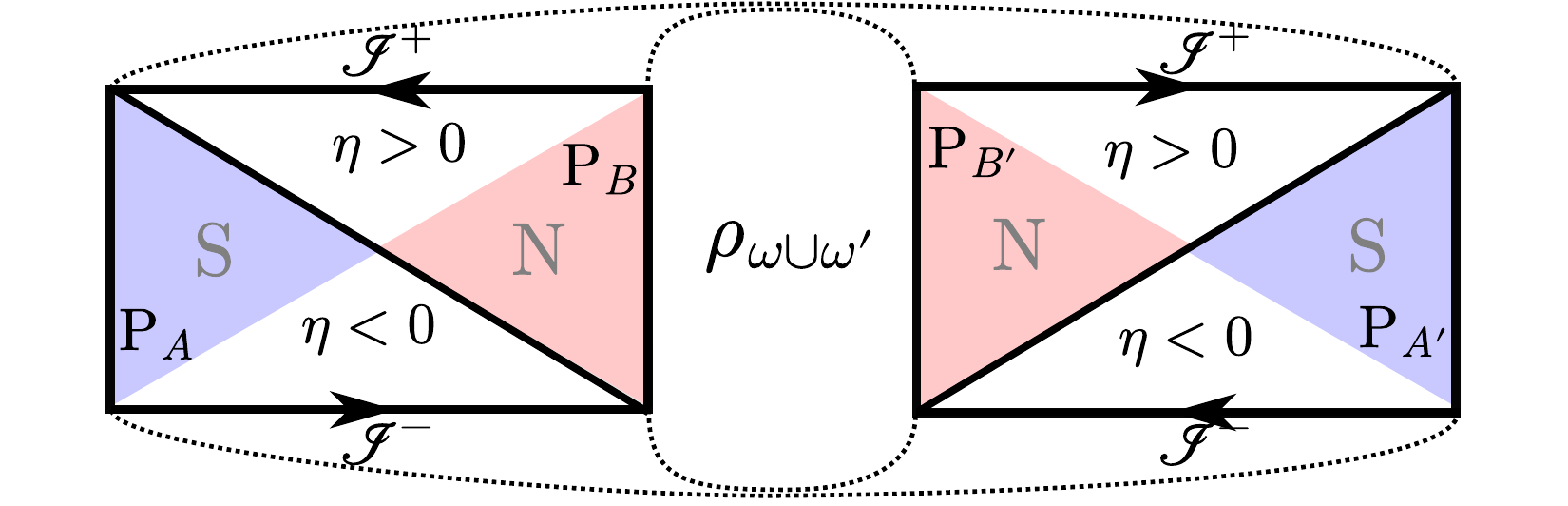}
  \caption{Penrose diagram of the $\nodstimesR$ quotient spacetime, composed of expanding and contracting Poincar\'e patches that are glued at their boundaries with a second pair of expanding and contracting Poincar\'e patches. A pair of ``north'' (`N') or ``south'' (`S') static patches support a vacuum that is conformally related to $\rho_{\omega \cup \omega'}$.}
  \label{fig:conformaldiagram}
\end{figure}

While the conformal extension doubles every point $(u_+,v_+,x_+,y_+)$ with $(u_-,v_-,x_-,y_-)$, the $\Sigma$ quotient identifies $(u_\pm,v_\pm,x_\pm,y_\pm) \rightarrow -(u_\pm,v_\pm,x_\pm,y_\pm)$. According to Eqs.~\ref{eq:dSglobal1}-\ref{eq:dSglobal4}, this takes one $\ds\times\reals$ to the other with $t \rightarrow -t$ and $\theta \rightarrow \pi - \theta$ in de Sitter global coordinates, and identifies one hyperbolic branch $\reals$ with another. As a result, the $\Sigma$ quotient effectively reduces the number of hyperbolic branches down to one. In summary, the metric of the resultant $\nodstimesR$ is simply the same as $\ds\times\reals$ glued to another along their conformal boundaries and at $\chi\rightarrow\pm\infty$ (see Fig.~\ref{fig:conformaldiagram}).

It remains to verify that the boundaries of $\nodstimesR$ qualify as asymptotically $\ds$ despite the extra fifth dimension and the gluing. To see whether the conformal suppression of the added fifth dimension is rapid enough not to affect the asymptotic symmetries, we can analytically continue the Brown-Henneaux boundary conditions from AdS$_4$ to de Sitter with global metric $\dd s^2 = \frac{\ell^2}{\cos^2 t} (-\dd t^2 + \dd \Omega_3^2)$. This reveals that, since the $\reals$ factor of $\nodstimesR$ is independent of $t$, the only Brown-Henneaux condition that needs to be verified is $g_{AB} = (t\mp\pi/2)^{-2} \gamma_{AB} + \mathcal O(1)$ where $\gamma_{AB}$ is the fixed metric on the spatial sphere at timelike infinity ($t \rightarrow \pm \pi/2$) and $x^A = (r, \Omega_2)$~\cite{Strominger01}. This condition is clearly met by both $\nodstimesR$ and regular $\ds \times \reals$.

\subsection{The Bulk/Boundary Correspondence}
\label{sec:dscftcorrespondence}

  We found in the previous section that the $\nodstimesR$ spacetime consists of two $\ds\times\reals$ spacetimes that are glued together antipodally on their similar conformal boundaries.

Since $\noadstwosphere$ and $\nodstimesR$ are derived by foliating the same $\mathbb R^{2,5}$ lightcone (i.e.~Eq.~\ref{eq:noadstwosphererestriction1} and Eq.~\ref{eq:dstimesRconstraint1}), where the $\RR{2}{2}$ part is extended, they contain conformally flat patches that are also conformal to each other~\cite{Leuven18}. Specifically, these conformally flat patches correspond to the Rindler wedge of $\ads \times \twosphere$ and the static patch of $\ds \times \reals$~\cite{Anninos12}.

The Rindler wedge of orientable or non-orientable $\ads \times \twosphere$ has metric,
\begin{equation}\label{eq:AdS-Rindler metric}
    ds^2 = -\left(\frac{r^2}{\ell^2}-1\right)\dd t^2 + \left(\frac{r^2}{\ell^2}-1\right)^{-1}\dd r^2 + r^2\dd\phi^2 + \ell^2 \dd\Omega_2^2,
  \end{equation}
  where $r \in (\ell,\infty)$, $t \in \mathbb R$, $\phi \in \reals$, and $\dd \Omega_2^2$ is the $\twosphere$ metric. The static patch metric of orientable $\ds \times \reals$ or $\nodstimesR$ is
\begin{equation}\label{eq:dS-static metric}
       ds^2 = -\left(1-\frac{R^2}{\ell^2}\right)\dd t^2 + \left(1-\frac{R^2}{\ell^2}\right)^{-1}\dd R^2 + R^2 \dd\Omega_2^2 + \ell^2\dd\phi^2.
\end{equation}
  where $R \in (0,\ell)$, $t \in \mathbb R$, $\phi \in \mathbb{R}$, and $\dd \Omega_2^2$ is the $\twosphere$ metric.
Multiplying Eq.~\ref{eq:dS-static metric} with $\ell/r=R/\ell$ conformally maps it to Eq.~\ref{eq:AdS-Rindler metric} (see Appendix~\ref{app:conformal_map} for more details). These patches can be fully contained in the Poincar\'e patches of either the orientable and non-orientable $\ads\times\twosphere$, or $\ds\times\reals$ and $\nodstimesR$. 

Since the conformal map is between a single Rindler wedge and static patch, in general it is only useful for conformally defining a static patch vacuum. Extending the conformal map to include the antipodal static patch is not possible. However, for $\noadstwosphere$ and $\nodstimesR$, it turns out to be possible to extend the conformal map in the other direction (i.e.~through the conformal boundary of $\noadstwosphere$) and include the Rindler wedge that shares its boundary with the first Rindler wedge. Heuristically, the shared Dirichlet condition on either side of the gluing surface requires that the induced boundary actions of any bulk conformal transformations on regions which share a boundary must be mutually compatible.
  
We conjecture that this compatibility can be seen through the higher-dimensional analogue of the Schwarz reflection principle~\cite{Schwarz1870,Davis58}, which originally specifies criteria under which the domain of analytic functions on manifolds with boundary can be extended beyond the boundary. We outline a generalization as follows. Consider again a pair of $\noads\times\twosphere$ Rindler wedges $\omega\,\cup\,\omega'$ sharing a boundary causal diamond $\partial\omega=\partial\omega' = \Omega$. A conformally compactified Rindler wedge is a pseudo-Riemannian manifold (with a class of conformally-equivalent metrics) with conformal boundary. We assume that the conformal transformation between the metrics in Eq.~\ref{eq:AdS-Rindler metric} and Eq.~\ref{eq:dS-static metric} can be completed to act on $\Omega$. Given such a conformal map $\Phi:\omega \rightarrow  \Phi(\omega)\equiv\omega_s$, where $\mathcal \omega_s$ is the static patch, since conformal transformation in greater than two-dimensions consist of M\"obius transformations and a local reflection across a hypersurface generalizes to a spherical inversion, we can define such an inversion, $\mathcal R_R(x)$, across the Rindler conformal boundary $\Omega$, and a similar inversion, $\mathcal R_S$, relating static patches on opposite sides of the $\nodstimesR$ gluing surface. The analogue of the Schwarz reflection principle to conformal transformations on higher-dimensional Riemannian manifolds suggests that $\mathcal R_R$ and $\mathcal R_S$ exist~\cite{Shapiro92}. We thus define the larger piece-wise map
  \begin{equation}
    \tilde \Phi(x) =
    \begin{cases}
      \Phi(x) & \text{for }x \in \omega\\
      \Phi'(x)\equiv \mathcal R_S \circ \Phi \circ \mathcal R_R(x) & \text{for }x \in \mathcal \omega'.
    \end{cases}
  \end{equation}
  The combined map $\tilde \Phi(x)$ is smooth across the conformal boundary $\Omega$, in the sense that the induced actions of $\Phi$ and $\Phi'$ on the shared boundary $\partial\omega=\partial \mathcal\omega'$ are identical.
  
It can also be seen that the conformal transformation $\tilde{\Phi}$ preserves the entanglement structure of the vacuum. The action of $\tilde\Phi$ is decomposed into the action of $\Phi$ and $\Phi^*$, simultaneously. These transformations induce, respectively, corresponding transformations of the field supported on the two regions $\omega$ and $\omega'$. Since the state of the Rindler wedges is of the form Eq.~\ref{eq:post-quotient}, which is supported on two non-interacting bulk Hilbert spaces, the induced action on the fields is of the form $U_{\Phi}\otimes U_{\Phi^*}$.   Since the operator $U_\Phi\otimes U_{\Phi'}$ factorizes, it does not change the entanglement structure of $\rho_{\omega\,\cup\,\omega'}$.  In particular, the entanglement between $\rho_\omega$ and $\rho_{\omega'}$ is identical to the entanglement between $\rho_{\Phi(\omega)}$ and $\rho_{\Phi'(\omega')}$.

The entanglement between $\rho_\omega$ and $\rho_{\omega'}$ is a consequence of how $\omega$ and $\omega'$ are embedded in the global $\noads\times\twosphere$ space, its topology. The same is true for $\Phi(\omega)$ and $\Phi(\omega')$, i.e, their entanglement reflects how they are embedded within $\nodstimesR$. In both cases, the wedges/patches are related to one other through bulk time evolution, and share points on the gluing surface at their boundaries. They intersect with each others' casual futures and pasts, suggesting a nontrivial entanglement between $\omega$ and $\omega'$ ($\omega_s$ and $\omega_s'$, resp.). Contrast this with a similar transformation of pairs of Rindler vacua before the quotient is taken. In that case, the state $\rho_\Omega$ factorizes across $\mathcal{H}_{CFT}\otimes\mathcal H_{CFT}$, and this is reflected in the bulk state Eq.~\ref{eq:pre-quotient}. For both $\rho_{\omega\,\cup\,\omega'}$ and $\rho_{\Phi(\omega)\,\cup\,\Phi'(\omega')}$, factorization is reflected in the causal structure of the spaces with which $\omega\,\cup\,\omega'$ and $\Phi(\omega)\cup\Phi'(\omega')$ are embedded: i.e. they are supported on non-interacting boundary-glued $\ads\times\twosphere$ spaces, or similarly non-interacting boundary-glued $\ds\times\R$ spaces.  In summary, the conformal vacuum on a pair of $\noads\times\twosphere$ Rindler wedges $\omega\,\cup\,\omega'$ is conformal to that of a pair of $\nodstimesR$ static patches $\Phi(\omega)\cup\Phi'(\omega')$ 

\begin{figure}
  \centering
  \includegraphics[width=1.0\linewidth]{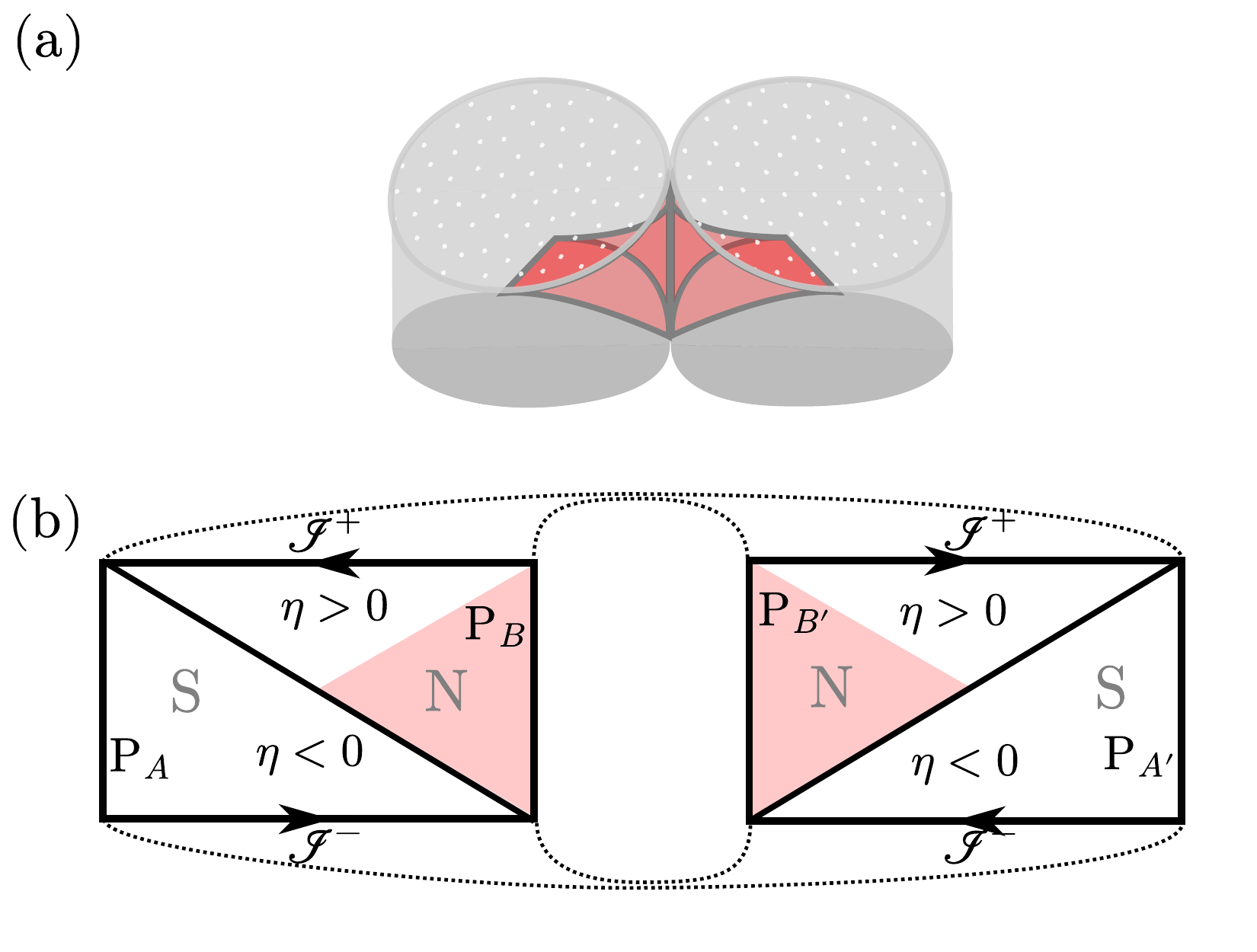}
  \caption{The conformal vacuum is supported on (a) Rindler wedges with shared boundary in $\noads\times\twosphere$ and (b) static patches with glued-together worldlines in $\nodstimesR$.
  }
  \label{fig:conformalvacuum}
\end{figure}

\section{Discussion}
\label{sec:discussion}

In Section~\ref{sec:noads}, we constructed $\noads$ as an orientation reversing $\mathbb Z_2$ quotient, $\Sigma$, of two vacuum $\ads$ spaces glued along their asymptotic boundaries. We demonstrated that the quotient acts on the $\onesphere\times\twosphere$ global topology to produce a twisted product topology $\onesphere 
\tildetimes \twosphere$, which is compact and non-orientable. We also demonstrated that the topology of the gluing surface along the shared (internal) boundary submanifold is an orientable twisted torus, which is homeomorphic to a torus with two independent null directions, $\onesphere\times\onesphere$. Such a topology is equivalently described by two Poincare patches glued along their shared asymptotic boundary, whose null directions have been compactified to $\onesphere\times\onesphere$, rather than the usual $\mathbb R\times\mathbb R$. The universal cover, however, is identical to that of two glued $\ads$ spaces, whose topology is $\R\times\twosphere$, and whose asymptotic boundary is an infinite (timelike) $\reals\times\onesphere$ cylinder.

As an aside, another way to describe $\noads$ is as a different compactification of a single-cover of $\ads$. In the more common case, the bulk is compactified on a (two-sided) Lorentzian torus, which is periodic in time and space with $2 \pi$ periods. In the compactified space, this torus is still two-sided. Since it is a true external manifold, only one side is relevant to the bulk. By comparison, the $\Sigma$ quotient construction is equivalent to a compactification of the bulk on a (two-sided) Lorentzian torus, which is periodic in two null directions with $\pi$ periods. In the compactified space, this torus is one-sided, since it is an internal manifold, and so both sides are relevant to the bulk; its normal line bundle is non-trivial. This means that, individually, the bulk and internal conformal boundary of $\noads$ are two-sided (i.e.~orientable). However, together as a conformally compactified manifold, the space is non-orientable.

We have argued that although the gluing surface is not an external boundary in the traditional sense, there is still a definable asymptotic region, so that the degrees of freedom at the boundary can be described by a holographic theory supported on a Lorentzian torus with null periodic lattice. States supported on this internal boundary submanifold consist of states on two CFT$_2$s, each of with naturally factorize into two independent chiral components, but that are in a permutation-twisted sector that prevents factorized (i.e.~separable) states on each CFT$_2$. Since their bulk duals consist of entangled states between bulk Hilbert space factors associated with opposite sides of the boundary, this agrees with the feature that their CFT duals are also entangled across their two factors. The entanglement in the bulk is a consequence of causal connectedness of the bulk, not a dynamical coupling, since we impose Dirichlet boundary conditions on the gluing surface. We also note, however, that because the boundary has two compact null directions, and the bulk has a compact time direction, both the bulk and the boundary necessarily contain closed causal curves. This may suggest instability under perturbations if the appropriate subspace is not restricted to, or other pathologies such as violations of the Null Energy Condition.

Using a similar method, we constructed the space $\noadstwosphere$, wherein two copies of $\ads\times\twosphere$ are glued and quotiented. It is well known that Rindler wedges of $\ads\times\twosphere$ are conformal to static patches of $\ds\times\R$. However, using a conjectured higher-dimensional generalization of the Schwarz reflection principle, we argued that that conformal transformations on Rindler wedges of $\noadstwosphere$, which share a conformal boundary, act simultaneously on both $\ads$ bulk components. This means that such simultaneous conformal transformations can relate boundary-glued pairs of Rindler wedges to pairs of static patches in separate $\ds\times\R$ spaces that are also glued together and share a worldline.

An important feature of the conformal relationship is that entanglement across the $\noadstwosphere$ boundary is mirrored by the entanglement across the $\nodstimesR$ static patches. In both cases, the non-trivial entanglement is a consequence of the fact that opposite sides of the boundary gluing surface are related to one another through the causal non-independence of the bulk on opposite sides. In the $\noadstwosphere$, the boundary itself is also restricted to a permutation-twisted sector that forbids separable states.

These results demonstrate the non-trivial holographic consequences that can result from topologically (but non-dynamically) gluing together two $\ads$ spacetimes and performing a subsequent quotient. Of course, adding in dynamic decoupling between the original two copies of $\ads\times\twosphere$ (or $\ds\times\reals$), would likely complicate matters further and is an interesting direction of future study. 

An important feature of $2+1$-gravity is that it is purely topological. A consequence of this is the existence of purely topological black holes, e.g. the BTZ black hole, whose thermal vacuum structure is more naturally analogous to that of the $\ds$ static patch vacuum than is the Rindler vacuum, due to its finite temperature. This raises the interesting question of whether constructions similar to the one outlined in this paper can be used to relate the conformal vacuum of non-orientable BTZ spaces with non-orientable $\ds$ spaces. We leave this investigation for future work.

\bigskip
\noindent\textbf{Acknowledgements}
\bigskip 

This material is based upon work supported by the U.S. Department of Energy, Office of Science, Office of Advanced Scientific Computing Research, under the Quantum Testbed Pathfinder program. Support is also acknowledged from the U.S. Department of Energy, Office of Science, National Quantum Information Science Research Centers, Quantum System Accelerator (Award No. DE-SCL0000121). SAND2026-28951O

Sandia National Laboratories is a multimission laboratory managed and operated by National Technology \& Engineering Solutions of Sandia, LLC, a wholly owned subsidiary of Honeywell International Inc., for the U.S. Department of Energy’s National Nuclear Security Administration under contract DE-NA0003525. This paper describes objective technical results and analysis. Any subjective views or opinions that might be expressed in the paper do not necessarily represent the views of the U.S. Department of Energy or the United States Government. 
  
\begin{appendix}

\section{Conformal Extension}
\label{app:Kleinextension}

  Consider the  metric of $\RR{2}{2}$ in Klein coordinates:
  \begin{align}
    \dd s^2 &= -\dd u^2 - \dd v^2 + \dd x^2 + \dd y^2\\
            &= -\dd q^2 - q^2 \dd \psi^2 + \dd r^2 + r^2 \dd \phi^2\\
            &= \Omega(U,V) \left( -\dd U \dd V - \frac{1}{4} \sin^2(V+U) \dd \psi^2 + \frac{1}{4} \sin^2(V-U) \dd \phi^2 \right),
  \end{align}
  where $q^2 = u^2 + v^2>0$, $r^2 = x^2 + y^2 >0$, $\tan U = q+r$, $\tan V = q-r$, $\Omega(U,V) = \cos^{-2}U \cos^{-2}V$. We then consider a conformal extension of the coordinates from $0 \le U < \pi/2$ and $|V| \le U$ to $0\le U < \pi$ and $0\le V < \pi$ with the conformal factor $\Omega(U,V)$ dropped (i.e.~a conformal compactification).

  Embedding $\ads$ into $\RR{2}{2}$ spacetime as $-u^2-v^2+x^2+y^2 = -\ell^2$, performing the conformal extension produces
\begin{equation}
  \dd s^2 = \ell^2 (-\dd t^2 + \dd \tilde \rho^2 + \sin^2 \rho \dd \phi^2),
\end{equation}
with $0 \le \rho < \pi$, where the conformal factor has been similarly dropped.

Consider the quotient of this doubled space by $\Sigma(U,V,\psi,\phi) \rightarrow (-U,-V,\psi+\pi,\phi+\pi)$. This is equivalent to $\Sigma(t,\rho, \phi) \rightarrow (t+\pi, \pi-\rho, \phi+\pi)$ on the embedded anti-de Sitter spacetimes. Since $(t,\phi) \rightarrow (t+\pi,\phi+\pi)$ takes Poincar\'e patch to its complementary patch, this is equivalent to gluing a Poincar\'e patch's conformal boundary to its complementary Poincar\'e patch's boundary with a twist. In terms of Poincar\'e coordinates with metric $\dd s^2 = (-\dd t^2 + \dd x^2 + \dd y^2)/y^2$, the twist can be expressed by $\dd y \rightarrow - \dd y$ at the conformal boundary $y\rightarrow\infty$. 

\section{$\noadstwosphere$ Construction}
\label{app:noadstwosphere}

To construct $\noadstwosphere$, we simply add a $\twosphere$ to $\noads$ as a trivial product space:
\begin{equation}
  \dd s^2_{\noadstwosphere}= \dd s^2_\noads + \ell^2 \dd \Omega_2^2,
\end{equation}
where $\dd \Omega^2$ defines the $\twosphere$. 

At the conformal boundary, the conformal factor makes the $\noads$ part of the metric infinitely larger than the $\twosphere$. Specifically, the Brown-Henneaux conditions for $\ads$ with global metric $\dd s^2 = -f(r) \dd t^2 + f^{-1}(r) \dd r^2 + r^2 \dd \theta^2$~\cite{Brown86} are:
 \begin{align}
   g_{AB} &= r^2 \gamma_{AB} + \mathcal O(1),\\
   g_{r A} &= \mathcal O (r^{-3}),\\
   g_{r r} &= \frac{\ell^2}{r^2} + \mathcal O (r^{-4}),
 \end{align}
where $\gamma_{AB}$ is a fixed metric on the cylinder at spatial infinity $r \rightarrow \infty$ and $x^A = (t, \theta)$. $\noadstwosphere$ satisfies these conditions since the $\twosphere$ bundle can be considered to add as a $\mathcal O(r^{-2})$ factor in $g_{AB}$, which is contained in $\mathcal O(1)$. 

\section{$\text{Rindler-AdS}_3\times \twosphere \cong \text{Static-dS}_4\times \mathbb{H}^1$ }
\label{app:conformal_map}
  
As demonstrated in \cite{Leuven18}, any two spacetimes which are conformally flat can be embedded in an $\mathbb{R}^{(2,D-2)}$ lightcone. Moreover, any two spacetimes which can be embedded in the same lightcone can be related to one another via a Weyl transformation. This can be used to establish conformal equivalences between many conformally flat spaces. Additionally, since local geometry is preserved under quotients, this method can also be used to establish the conformal equivalence of quotients of the embedded spaces. We first use this embedding space formalism to demonstrate the known conformal relationship between Rindler wedges of $\ads\times\twosphere$ and the static patches of $\ds\times\reals$ . We then establish a conformal relationship between quotients of their double covers, namely those which produce $\noads\times\twosphere$ and $\nodstimesR$. 

We consider a lightcone through the origin of $\mathbb{R}^{2,5}$ given by 
\begin{equation}
    -u^2 -v^2 + x^2 + y^2 + a^2 + b^2 + c^2 = 0
\end{equation}
$\ads\times\twosphere$ is obtained by the additional constraints 
\begin{align}
    \underbrace{-u^2 +x^2-v^2 + y^2}_{= - \ell^2} + \underbrace{a^2 + b^2 + c^2}_{=\ell^2} = 0.
\end{align}
We introduce an additional constraint to parameterize the AdS Rindler wedges:  
\begin{align}\label{eq:AdS-lightcone}
    \underbrace{-u^2 +x^2}_{=r^2 - \ell^2} +\underbrace{ -v^2 + y^2}_{= - r^2} + \underbrace{a^2 + b^2 + c^2}_{=\ell^2} = 0,
\end{align}
where $r$ is the Rindler radial coordinate. The induced metric $\dd s^2$ on this section of the lightcone is
\begin{equation}\label{eq:AdS-Rindler metric_app}
    \dd s^2 = -\left(\frac{r^2}{l^2}-1\right)\dd t^2 + \left(\frac{r^2}{l^2}-1\right)^{-1}\dd r^2 + r^2\dd\phi^2 + l^2 \dd\Omega_2^2.
\end{equation}

Similarly, $\ds\times\reals$ is embedded into the lightcone with a set of constraints: 
\begin{align}
    \underbrace{-u^2 +x^2+a^2+b^2 + c^2}_{= \ell^2} \underbrace{-v^2 + y^2}_{=-\ell^2} = 0
\end{align}
To parameterize the static patches, we introduce the additional constraint
\begin{align}\label{eq:dS-lightcone}
    \underbrace{-u^2 +x^2}_{=-R^2+\ell^2}+\underbrace{a^2+b^2 + c^2}_{=R^2} \underbrace{-v^2 + y^2}_{=-\ell^2} = 0
\end{align}
where $R$ is the static patch radial coordinate. The induced metric $\dd\tilde{s}^2$ is
\begin{equation}\label{eq:dS-static metric_app}
       \dd \tilde s^2 = -\left(1-\frac{R^2}{\ell^2}\right)\dd t^2 + \left(1-\frac{R^2}{\ell^2}\right)^{-1}\dd R^2 + R^2 \dd\Omega_2^2 + \ell^2\dd\phi^2.
\end{equation}
It is clear from \eqref{eq:AdS-lightcone} and \eqref{eq:dS-lightcone} that the set of constraints are identical, with the roles of the radial coordinates inverted, i.e. $(r/\ell)\leftrightarrow(R/\ell)$. Indeed, we also see that metrics \eqref{eq:AdS-Rindler metric_app} and \eqref{eq:dS-static metric_app} are related by a Weyl rescaling $\Omega^2 = \left(\frac{\ell}{r}\right)^2 = \left(\frac{R}{\ell}\right)^2$. In  particular,

\begin{equation}\dd \tilde{s}^2 = \left(\frac{l}{r}\right)^2\cdot \dd s^2 \Big\vert_{\substack{r/\ell=\ell/R}}
\end{equation}

Finally, since the induced metrics on the Rindler wedges and static patches of $\ads\times\twosphere$ and $\ds\times\R$ (resp.)~are locally identical to those on their non-orientable counterparts, $\noadstwosphere$ and $\nodstimesR$, the same conformal relationship holds for them as well.  

\end{appendix}

\bibliographystyle{unsrt}
\bibliography{biblio}

\end{document}